\documentclass[journal,letterpaper,twoside,10 pt]{IEEEtran}

\usepackage{cite}
\usepackage[colorlinks = true, urlcolor = blue,citecolor = black, linkcolor = black]{hyperref}
\usepackage{balance}
\usepackage[font = footnotesize]{caption}
\usepackage{amssymb}
\usepackage{mathtools}
\usepackage{amsmath}
\usepackage{url}
\usepackage{optidef}
\usepackage{subcaption}
\usepackage{graphicx}
\usepackage{tikz}

\newtheorem{theorem}{Theorem}[section]

\newtheorem{problem}[theorem]{Problem}

\newtheorem{definition}[theorem]{Definition}

\begin{document}

\title{ Distributed Perception Aware Safe Leader Follower System via Control Barrier Methods}

\author{Richie R. Suganda$^{1}$, Tony Tran$^{3}$, Miao Pan$^{1}$, Lei Fan$^{2}$, Qin Lin$^{2}$ and Bin Hu$^{2}$%
\thanks{$^{1}$Richie R. Suganda and Miao Pan are with the Department of Electrical and Computer Engineering at University of Houston, Houston, TX 77004, USA. (e-mail: {\tt rrsugand@cougarnet.uh.edu; mpan2@central.uh.edu}).}
\thanks{$^{2}$ Lei Fan, Qin Lin, and Bin Hu are with the Department of Engineering Technology at University of Houston, Houston, TX 77004, USA. (e-mail: {\tt lfan8@central.uh.edu; qlin21@central.uh.edu; bhu11@central.uh.edu}).}
\thanks{$^{3}$ Tony Tran is with Cullen College of Engineering Research Computing at University of Houston, Houston, TX 77004, USA. (e-mail: {\tt thtran37@cougarnet.uh.edu)}.}

\thanks{This work was supported by NSF CHS Award $2007386$.  }
}

\maketitle

\begin{abstract}
This paper addresses a distributed leader-follower formation control problem for a group of agents, each using a body-fixed camera with a limited field of view~(FOV) for state estimation. The main challenge arises from the need to coordinate the agents' movements with their cameras' FOV to maintain visibility of the leader for accurate and reliable state estimation. To address this challenge, we propose a novel perception-aware distributed leader-follower safe control scheme that incorporates FOV limits as state constraints. A Control Barrier Function~(CBF) based quadratic program is employed to ensure the forward invariance of a safety set defined by these constraints. Furthermore, new neural network based and double bounding boxes based estimators, combined with temporal filters, are developed to estimate system states directly from real-time image data, providing consistent performance across various environments. Comparison results in the Gazebo simulator demonstrate the effectiveness and robustness of the proposed framework in two distinct environments. 

\end{abstract}

\section{Introduction}
\label{sec:intro}

\begin{figure}[t]
	\begin{center}
		\medskip
		\includegraphics[width=\linewidth]{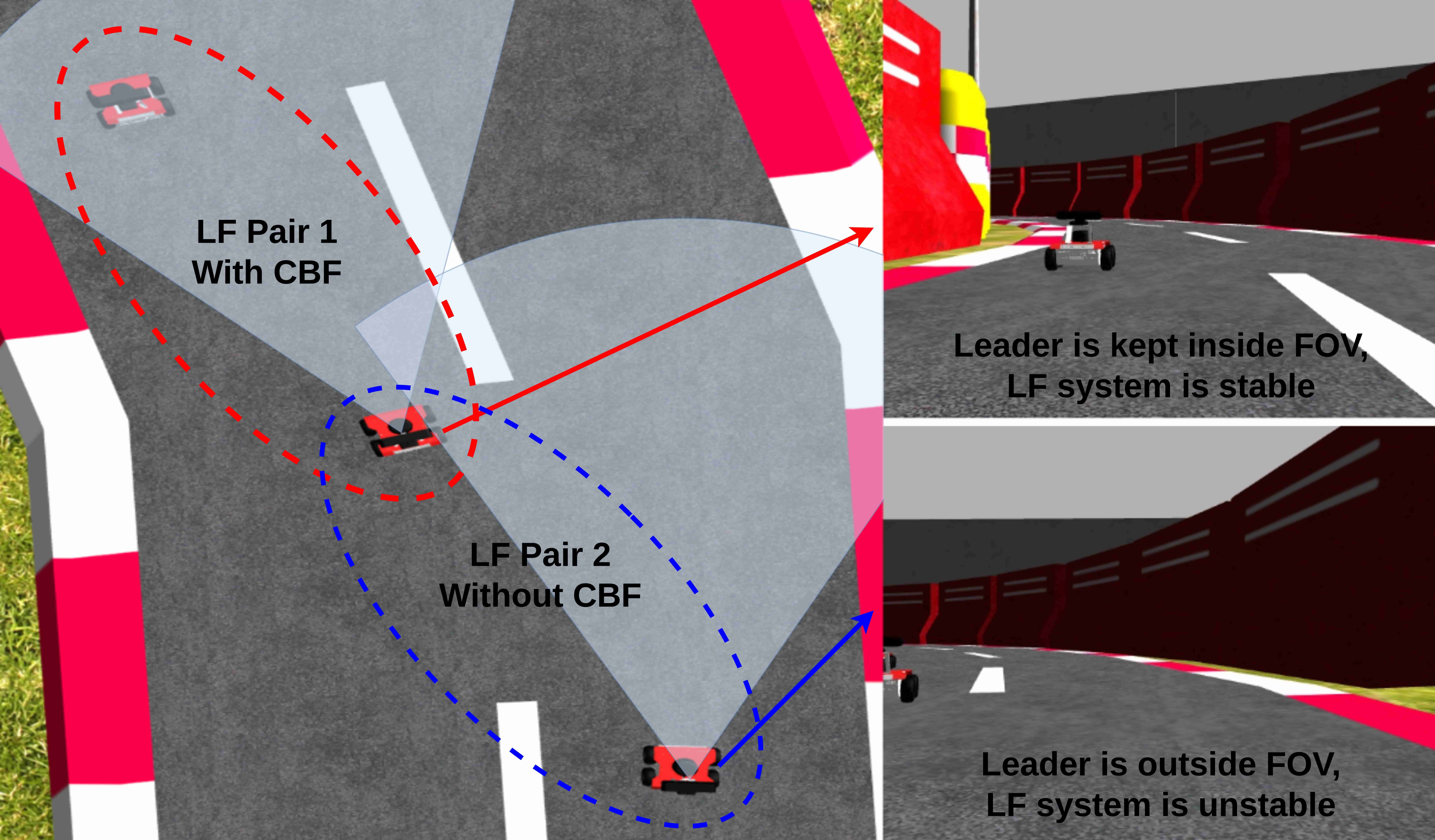}
	\caption{\textbf{Perception-aware Leader-follower (LF) System with Limited FOV.} LF formations are composed of robots equipped with onboard body-fixed camera that have limited FOV. To maintain safe and reliable formations, the leader must remain within the follower's FOV.}
		\label{fig:motivation}
	\end{center}
	\vspace{-4ex}
\end{figure}

\IEEEPARstart{L}{eader}-follower formation control remains a widely adopted approach to accomplish collective tasks among multiple agents, such as environmental monitoring \cite{Panagou2014}, search and rescue\cite{Gomez2022}, and surveillance \cite{Cui2010}. However, achieving effective and safe formation control in real-world scenarios is challenging particularly when agents cannot access GPS signals and can only rely on limited onboard sensing capabilities, such as body-fixed cameras with restricted FOV. The need to keep the leader within the camera’s limited FOV to ensure accurate state estimation creates a dependency between the agents’ movements and their perception capabilities. This dependency imposes constraints on the motion of follower agents, complicating the control task as shown in Fig.~\ref{fig:motivation}. To address such challenges, this paper develops a safe distributed leader-follower formation control problem where each agent uses a body-fixed camera for state estimation.

Estimating leader-follower system states using cameras has been explored in numerous studies, which can be mainly categorized as marker-based and deep learning-based methods. Early works, such as \cite{garrido2014automatic, fiala2005artag, olson2011apriltag, mariottini2009vision,Panagou2014}, relied on markers with easily detectable features to estimate vehicle positions. However, these methods have limitations in real-world situations, as they require markers to be placed on objects, depend on known camera characteristics, and are sensitive to environmental factors such as marker wear and tear. More recent studies have utilized deep neural networks (DNNs) and reinforcement learning to estimate states or control systems by learning the mapping between images and system states or control actions \cite{Khodamipour2023,Zhou2019,Wang2016}. While the estimators proposed in this paper align with these deep learning methods, they combine a novel temporal filter with the neural network model and double bounding box based method to capture the spatiotemporal features present in images of moving leaders, an aspect that has not been adequately addressed in existing works.


Control Barrier Functions (CBFs) have been extensively studied and applied in various safety-critical applications, including lane keeping \cite{Ames2016, ames2019control}, bipedal locomotion \cite{Hsu2015, Agrawal2017}, and obstacle avoidance \cite{Mestres2024, chen2020guaranteed, wang2016safety}, as an effective approach to balance control and safety requirements. Recent efforts have extended CBFs to develop perception-aware safe robotic systems, focusing on obstacle avoidance \cite{bena2023safety, salehi2021constrained, Mestres2024, DeCarli2024} and object tracking \cite{Ning2024, salehi2021constrained, Balandi2023}. For example, \cite{bena2023safety} employs CBFs to optimize the camera direction for collision avoidance, while \cite{salehi2021constrained} uses CBFs in visual servoing to control a robot arm, ensuring the target remains within the camera's FOV. In \cite{Mestres2024}, a distributed CBF is developed to ensure safe navigation for multi-agent systems, assuming each robot has perfect knowledge of its neighbors' states.

However, the perception-aware methods above cannot be directly applied to the leader-follower system considered in this paper. Prior works either overlook the state estimation challenges posed by image-based data or fail to account for the intrinsic coupling between agents' movements and their perception limitations - an essential factor for ensuring safe leader-follower control. This paper addresses these gaps by proposing a distributed, perception-aware, and safe leader-follower formation control and estimation framework.


Our contributions are threefold. First, we propose a novel perception-aware, distributed leader-follower control scheme that explicitly incorporates camera FOV limitations as state constraints. A CBF-QP framework is employed to enforce forward invariance of a safety set defined by these constraints. Second, new neural network based and double bounding boxes based estimators combined with temporal filters are introduced to obtain reliable estimates for leader-follower states across different environments. Third, extensive comparison results in the Gazebo simulator are provided to validate the proposed approaches. 

The rest of the paper is structured as follows: Section \ref{sec:prelim} provides the preliminaries. Section \ref{sec:sys} introduces the framework for leader-follower systems and problem formulation; Section \ref{sec:main-results} discusses perception aware distributed safe formation control scheme; Section \ref{sec:estimation} focuses on system state estimation; Section \ref{sec:experiment} presents experimental results in Gazebo simulator, and Section \ref{sec:conclusion} concludes the paper.
\section{Preliminaries}
\label{sec:prelim}
The fundamental notations and preliminaries on CBFs presented in this section are based on the work in \cite{Ames2016}.
 
\emph{Notation:} Let $\mathbb{R}, \mathbb{R}_{\geq 0}$ and $\mathbb{R}_{+}$ denote the set of real, non-negative and positive real numbers, respectively. Let $[n]=\{1, 2, \ldots, n\}$ denote a finite set of natural numbers. Let $\mathbb{R}^{n}$ denote a $n$ dimensional Euclidean vector space, and the infinity norm of a real vector $x \in \mathbb{R}^{n}$ is denoted by $|x| \triangleq \max_{1 \leq i \leq n} |x_i|$ with $x_{i} \in \mathbb{R}$ representing the $i^{th}$ element in the vector $x$. A real-valued function $f: \mathbb{R}^{n} \rightarrow \mathbb{R}^{m}$ is locally Lipschitz over a compact set $\mathcal{C} \subset \mathbb{R}^{n}$ if $\forall x, x' \in \mathcal{C}$, there exists a positive constant $M \in \mathbb{R}_{+}$ such that $|f(x) - f(x')| \leq M|x - x'|$. A real positive function $\beta: [0, a) \rightarrow \mathbb{R}_{\geq 0}$ is a class $\mathcal{K}$ function for $a \in \mathbb{R}_{+}$ if it is continuous and strictly increasing, and $\beta(0) = 0$. A continuous function $\alpha: (-b, a) \rightarrow (-\infty, \infty)$ with $a, b \in \mathbb{R}_{+}$ belongs to \emph{extended} class $\mathcal{K}$ if it is strictly increasing and $\alpha(0) = 0$ (Definition 2 in \cite{Ames2016}). It is clear that the extended class $\mathcal{K}$ functions generalize the class $\mathcal{K}$ to include bounded negative domains.

Consider an affine control system defined as below 
\begin{align}
	\label{eq:sys}
	\dot{x} = f(x) + g(x)u
\end{align}
where $x \in \mathbb{R}^{n}$ and $u \in \mathcal{U} \subset \mathbb{R}^{n}$ are system states and control inputs that are confined to a compact set $\mathcal{U}$. The functions $f$ and $g$ are assumed to be locally Lipschitz with appropriate dimensions. Given a control system defined in \eqref{eq:sys}, the safety specifications are enforced by ensuring system trajectories $x(t) \in \mathbb{R}^{n}$ are forward invariant over a closed set $\mathcal{C} \in \mathbb{R}^{n}$. In particular, the set $\mathcal{C}$ is defined by inequalities as $\mathcal{C} = \{x \in \mathbb{R}^{n} : h(x) \geq 0\}$ with $h: \mathbb{R}^{n} \rightarrow \mathbb{R}$ being a continuously differentiable function. 
With a non-empty closed set $\mathcal{C}$ defined above, a (Zeroing)~Control Barrier Function~(ZCBF) is introduced to ensure that the set $\mathcal{C}$ is forward invariant under the system dynamics in \eqref{eq:sys}.
\begin{definition}(ZCBF in \cite{Ames2016})
	\label{def:cbf}
For a given set $\mathcal{C} \subset \mathbb{R}^{n}$ defined by a continuously differentiable function $h:\mathbb{R}^{n} \rightarrow \mathbb{R}$, the function $h$ is called a ZCBF defined on a set $\mathcal{D} \subseteq \mathcal{C}$, if there exists an extended class $\mathcal{K}$ function $\alpha$ such that 
\begin{align}
	\label{ineq:cbf}
	\sup_{u \in \mathcal{U}}\{ L_f h(x) + L_g h(x)u + \alpha(h(x))\} \geq 0, \forall x \in \mathcal{D}.
\end{align}
where $L_f h(x) = \nabla h(x)f(x)$ and $L_gh(x) = \nabla h(x)g(x)$.
\end{definition}
With the condition \eqref{ineq:cbf}, one can further show that the set $\mathcal{C}$ is rendered forward invariant if control actions $u \in \mathcal{U}$ can be found to enforce the condition \eqref{ineq:cbf}. 
	
\section{System Framework and Problem Formulation}
\label{sec:sys}
\begin{figure}[t]
	\begin{center}
		\medskip
		\includegraphics[width=0.9\linewidth]{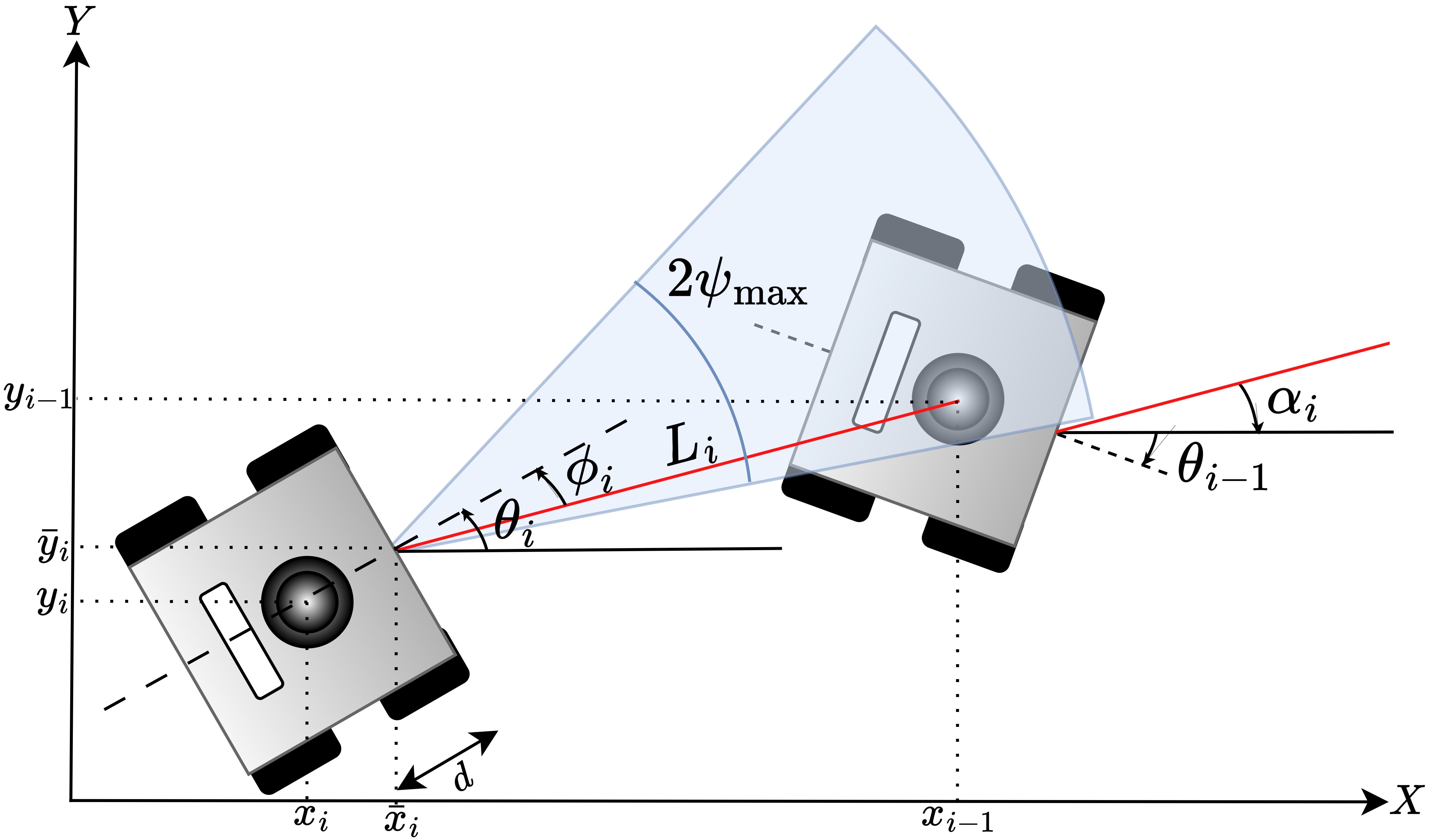}
		\caption{{Leader-follower formation diagram.}}
		\label{fig:lead-fol}
	\end{center}
	\vspace{-4ex}
\end{figure}
This paper considers a leader-follower formation system in a group of $n$ agents, where the kinematic of each agent $i$ is modeled as below,
\begin{align}
\dot{x}_{i} = v_{i}\cos{\theta_i}, \quad \dot{y}_{i} = v_{i} \sin{\theta_i},  \quad \dot{\theta}_{i} = \omega_{i}
\label{sys:agent}
\end{align}
where $x_{i}, y_{i}, \theta_{i} \in \mathbb{R}$ represent the horizontal and vertical coordinates of the agent's center and the bearing angle defined in the global frame, respectively. The control inputs $v_{i}$ and $\omega_{i}$ correspond to the agent's linear and angular velocities. A leader-follower formation system with $n$ agents consists of $n-1$ leader-follower pairs, as illustrated in Fig.~\ref{fig:lead-fol}.

The control objective for the leader-follower system is for each follower~(agent $i$) in the leader-follower pair to regulate its control inputs $v_{i}$ and $\omega_{i}$, in order to maintain a specified inter-vehicle distance $L_i$ and bearing angle $\alpha_i$ relative to its immediate leader (agent $i-1$). As shown in Fig.~\ref{fig:lead-fol}, $L_i$ represents the Euclidean distance between the follower's front end and the leader's center, while $\alpha_i$ denotes the relative bearing angle of the leader's heading with respect to the distance vector. Given the displacement $d$ from the agent’s center to its front, the global coordinates of the front midpoint are $\bar{x}_{i} = x_{i} + d\cos{\theta_i}$ and $\bar{y}_{i} = y_{i} + d\sin{\theta_i}$. The inter-vehicle distance $L_i$ is defined as follows: 
\begin{align}
L_{i} = \sqrt{(x_{i-1} - \bar{x}_{i})^2 + (y_{i-1} - \bar{y}_{i})^2}
\end{align}
Similar to $\alpha_i$ which defines the leader's relative bearing angle, $\phi_{i}$ represents the follower's relative bearing angle with respect to the distance vector $L_i$, as shown in Fig.~\ref{fig:lead-fol}. The geometric relationship between the angles $\phi_{i}, \theta_{i-1}, \theta_{i}$ and $\alpha_{i}$  is given by $\alpha_{i} = \theta_{i-1} - \theta_{i} - \phi_{i}$.  Building on the work presented in \cite{hu2015distributed,hu2021stochastic,mariottini2009vision}, the dynamics of the leader-follower pair, characterized by $L_i$ and $\alpha_i$, is modeled as follows:
\begin{align}
\resizebox{0.91\columnwidth}{!}{$
	   \begin{bmatrix}
		\dot{L}_i \\
		\dot{\alpha}_i
	\end{bmatrix}
	= 
	\underbrace{\begin{bmatrix}
		-\cos{\phi_{i}} & -d \sin{\phi_{i}}\\
		\frac{-\sin{\phi_{i}}}{L_i} & \frac{d\cos{\phi_{i}}}{L_i}
	\end{bmatrix}}_{g(\phi_i, L_i)}
		\underbrace{\begin{bmatrix}
		v_{i}\\
		\omega_{i}
	\end{bmatrix}}_{u_i} 
	+
\underbrace{\begin{bmatrix}
		\cos{\alpha_i} & 0\\
		\frac{-\sin{\alpha_i}}{L_i} & 1
	\end{bmatrix}}_{f(\alpha_i, L_i)}
	\underbrace{\begin{bmatrix}
		v_{i - 1}\\
		\omega_{i - 1}
	\end{bmatrix}}_{u_{i-1}}
$}
	\label{eq:leader-follower-pair}
\end{align}
where $v_{i-1}$ and $\omega_{i-1}$ represent linear and angular velocities of the leader in leader-follower pair $i$.

Given the dynamics in \eqref{eq:leader-follower-pair}, each follower in the leader-follower system is assumed to use an onboard camera with a limited FOV to measure and estimate the inter-vehicle distance, $L_i$, and the bearing angle, $\phi_i$. The orientation of each agent, $\theta_i$, is reliably estimated using its own IMU module, as described in \cite{Madgwick2011}. With $\phi_i$ and $\theta_i$, the follower computes the bearing angle, $\alpha_i$, using the equation $\alpha_i = \theta_{i-1} - \theta_{i} - \phi_{i}$.  The problem formulation is formally stated below, 
\begin{problem}
\label{problem}
Consider a leader-follower system with dynamics defined in \eqref{eq:leader-follower-pair}, let $\{[L_{i, d}; \alpha_{i, d}]\}_{i \in [n]}$ denote desired setpoints for each leader-follower pair, and $\mathcal{C}_{t}^{i}, \forall i \in [n]$ represent compact sets characterizing camera's coverage area for agent $i$ at time $t \in \mathbb{R}_{\geq 0}$. For a compact set  $\mathcal{U} \in \mathbb{R}^{m}$, the objective is to find distributed control laws $\mathcal{F}_{i} \subseteq \mathcal{U}, \forall i \in [n]$, that enable the system to achieve and maintain the desired formations specified by $\{[L_{i, d}; \alpha_{i, d}]\}_{i \in [n]}$ safely, while ensuring the leader remains within the follower's camera FOV, $\mathcal{C}_{t}^{i}$ at all times. 
\end{problem}

\begin{figure*}[t]
	\begin{center}
		\medskip
		\includegraphics[width=\linewidth]{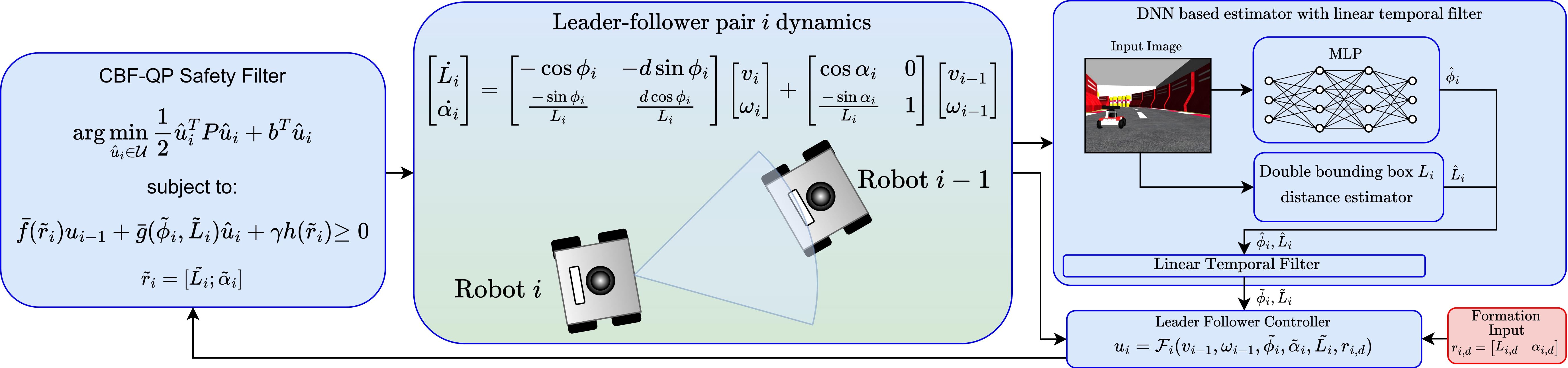}
		\caption{{Distributed Perception-aware Safe Leader-follower Control and Estimation Architecture.}
			\label{fig:distributed-safe-leader-follower}}
	\end{center}
	\vspace{-2ex}
\end{figure*}

To address Problem \ref{problem}, we develop a distributed perception-aware safe leader-follower control and estimation scheme, as illustrated in Fig.~\ref{fig:distributed-safe-leader-follower}. The following sections will provide detailed descriptions of each component of the proposed architecture. 


\section{Distributed Perception-aware Safe Leader-follower Formation Control}
\label{sec:main-results}
This section presents a perception-aware distributed formation control scheme that balances the control and safety objectives for a leader-follower system using CBF methods. 

Using the dynamics of leader-follower pair described in \eqref{eq:leader-follower-pair}, the distributed formation controller is designed through state feedback linearization techniques\cite{hu2021stochastic,Isidori1985}, as follows:
\begin{align}
u_{i}	=	\begin{bmatrix}
			v_{i}\\
			\omega_{i}
		\end{bmatrix}
		=&
		\begin{bmatrix}
			-\cos{\phi_{i}} & -L_i \sin{\phi_{i}}\\
			\frac{-\sin{\phi_{i}}}{d} & \frac{L_i\cos{\phi_{i}}}{d}
		\end{bmatrix}
		\biggl(
		\begin{bmatrix}
			K_{L_i}(L_{i,d}-L_i)\\
			K_{\alpha_i}(\alpha_{i,d}-\alpha_i)
		\end{bmatrix} \nonumber \\
		& \quad-
		\begin{bmatrix}
			\cos{\alpha_i} & 0\\
			\frac{-\sin{\alpha_i}}{L_i} & 1
		\end{bmatrix}
	 	\begin{bmatrix}
			v_{i - 1}\\
			\omega_{i - 1}
		\end{bmatrix}
		\biggl) \nonumber\\
		\triangleq&
		\mathcal{F}_{i}(v_{i - 1}, \omega_{i - 1}, \phi_{i}, \alpha_{i}, L_{i}, r_{i, d})
		\label{eq:controller}
\end{align}
where $K_{L_i}, K_{\alpha_i} \in \mathbb{R}_{>0}$ are linear controller gains for tracking desired distance and angle respectively. To compute the control actions $v_{i}$ and $\omega_{i}$, we assume that the follower's controller has access to the leader's linear and angular velocities $v_{i-1}$ and $\omega_{i-1}$ via wireless communication \cite{hu2015distributed}. 

With the distributed control law defined in \eqref{eq:controller}, it can be shown that the inter-vehicle distance $L_i$ and bearing angle $\alpha_i$ of each leader-follower pair will exponentially converge to the desired setpoints $L_{i, d}$ and $\alpha_{i, d}$, provided that the states, $\phi_{i}, \alpha_{i}, L_{i}$ obtained via on-board camera, are consistently available to the controller. However, this requirement may not always be met due to the coupling between the camera's FOV and agents' movements. Therefore, it is crucial to account for these correlations in the controller design. 

This paper models the coverage areas of a camera with limited FOV as a circular sector, defined by its radius and central angle. We assume that the camera is mounted at the center of the vehicle's front end, a position that remains equivalent to other mounting locations through a translation and rotation transformation $T \in \mathsf{SE}(2)$. Additionally, we assume the FOV aligns with the direction of the inter-vehicle distance $L_i$ and bearing angle $\phi_i$. The FOV coverage sector forms a ``safe" region $\mathcal{C}_{t}^{i}$. To ensure the leader is within the follower's camera view, the inter-vehicle distance $L_{i}$ and bearing angle $\phi_i$ must be within this region. 

Formally, let $2\psi_{\text{max}}$ represent the total angular width of the camera's FOV. As shown in Fig.~\ref{fig:lead-fol}, the FOV extends from $-\psi_{\text{max}}$ to $\psi_{\text{max}}$, where $\psi_{\text{max}}$ is the farthest angle visible to the left, and $-\psi_{\text{max}}$ is the farthest angle visible to the right of the agent's heading. Similarly, let $[D_{\text{min}}, D_{\text{max}}]$ denote the depth range of the camera where $D_{\text{min}},  D_{\text{max}}\in \mathbb{R}_{>0}$ are the minimum and maximum distances a depth camera can operate. Since the camera is aligned with the agent's heading, the safe region $\mathcal{C}^{i}$ is defined in the local frame of the agent as below, 
\begin{align}
	\mathcal{C}^{i} \triangleq \left \{\begin{bmatrix}
		L_i \\
		\phi_i 
	\end{bmatrix} \in \mathbb{R}^{2} \bigg| \begin{bmatrix}
	D_{\text{min}} \\
	-\psi_{\text{max}}
	\end{bmatrix} \leq \begin{bmatrix}
	L_i \\
	\phi_i 
	\end{bmatrix} \leq \begin{bmatrix}
	D_{\text{max}} \\
	\psi_{\text{max}}
	\end{bmatrix} \right\}
	\label{set:safety}
\end{align}

With $\phi_i = \theta_{i-1} - \theta_{i} - \alpha_i$, the safety set described in \eqref{set:safety} can be reformulated using four CBFs as follows, 
\begin{align}
\mathcal{C}^{i} \triangleq
\begin{cases}
	h_{1}(L_i) = L_i - D_{min} \geq 0 \\
	h_{2}(L_i) = -L_i + D_{max} \geq 0 \\
	h_{3}(\alpha_i) = \theta_{i-1} - \theta_{i} - \alpha_i + \psi_{max} \geq 0 \\
	h_{4}(\alpha_i) = -\theta_{i-1} + \theta_{i} + \alpha_i + \psi_{max} \geq 0
\end{cases}	
\label{set:cbf}
\end{align}


Let $h = [h_1; h_2; h_3; h_4]$ and $r_i = [L_i ; \alpha_i]$, the CBFs can be written in a compact form 
\begin{align}
	h(r_i) = Ar_i + B + C(\theta_{i}-\theta_{i-1})
		\label{eq:cbf}
\end{align}
with 
\begin{align}
A= \begin{bmatrix}
	1 & 0 \\
	-1 & 0 \\
	0 & -1 \\
	0 & 1
\end{bmatrix},  B= \begin{bmatrix}
-D_{\text{min}}\\
D_{\text{max}} \\
\psi_{\text{max}} \\
\psi_{\text{max}}
\end{bmatrix}, C= \begin{bmatrix}
0\\
0 \\
-1 \\
1
\end{bmatrix}
\end{align}
Based on Definition \ref{def:cbf}, let the extended class $\mathcal{K}$ functions $\alpha(h_i) = \gamma_i h_i, \forall i=1,2,3,4$ be linear where $\gamma_i \in \mathbb{R}_{>0}, \forall i$, and $\gamma = \text{diag}(\gamma_1,\ldots, \gamma_4)$ is a real positive definite diagonal matrix. Given the CBF representation in \eqref{eq:cbf} and the dynamics of leader-follower pair in \eqref{eq:leader-follower-pair}, the conditions for ensuring that the safety set $\mathcal{C}^{i}$ remains forward invariant are given by
\begin{align}
Af(\alpha_i, L_i)u_{i-1} + Ag(\phi_i, L_i)u_i + C(\omega_{i}- \omega_{i-1}) + \gamma h(r_i) \geq 0.
\label{ineq:forward-inv}
\end{align}
Given
\begin{align*}
	C(\omega_{i}- \omega_{i-1}) &= C\left(\begin{bmatrix}
		0 & 1
	\end{bmatrix}\begin{bmatrix}
		v_{i} \\
		\omega_{i}
	\end{bmatrix} -\begin{bmatrix}
	0 & 1
	\end{bmatrix}\begin{bmatrix}
	v_{i-1} \\
	\omega_{i-1}
	\end{bmatrix}\right) \\
	&= \bar{C}(u_i - u_{i-1})
\end{align*}
with $\bar{C} = C[0, 1]$, the CBF condition in \eqref{ineq:forward-inv} is rewritten as 
\begin{align}
	[Af(r_i)-\bar{C}]u_{i-1} +[Ag(\phi_i, L_i)+\bar{C}]u_i +\gamma h(r_i) \geq 0 
	\label{ineq:cbf-compact}
\end{align}
where the CBF conditions in \eqref{ineq:cbf-compact} are affine in  follower's control input $u_{i}$, allowing them to be formulated as a QP. 

The conditions in \eqref{ineq:cbf-compact} act as a safety filter for the control actions produced by the leader-follower formation controller defined in \eqref{eq:controller}. If the inequality in \eqref{ineq:cbf-compact} is violated due to the control input $u_i$ from \eqref{eq:controller}, we must adjust the control input to satisfy the safety constraints while minimizing deviations from the formation control objective. This adjustment strategy, based on the seminal work in \cite{Ames2016}, is formulated as a QP of the following form, 

\noindent\rule{\linewidth}{0.5pt}
\noindent \textbf{CBF-QP}
\begin{argmini}|l|
	{\hat{u}_i \in \mathcal{U}}{\frac{1}{2}(\hat{u}_i - u_i)^TP(\hat{u}_i - u_i) = \frac{1}{2}\hat{u}_{i}^{T}P\hat{u}_{i} +b^{T}\hat{u}_i }{}{}    
	\addConstraint{\bar{f}(r_i)u_{i-1} +\bar{g}(\phi_i, L_i)\hat{u}_i +\gamma h(r_i)}{\geq 0}    \nonumber
\end{argmini}
\noindent\rule{\linewidth}{0.5pt}
where $b = -P^{T}u_{i}$, $\bar{f}(r_i)= Af(r_i)-\bar{C}$ and $\bar{g}(\phi_i, L_i)= Ag(\phi_i, L_i)+\bar{C}$. To verify the CBF condition in \eqref{ineq:cbf-compact} and solve the QP, the follower requires information on $L_i, \phi_{i}$ as well as $u_{i-1}$ and $\theta_{i-1}$. This paper assumes that $u_{i-1}$ and $\theta_{i-1}$ are communicated to the follower via wireless channels. The next section introduces novel estimation methods to obtain the distance $L_i$ and $\phi_{i}$ using the follower's onboard camera. 
\section{Leader-follower System State Estimation}
\label{sec:estimation}
\begin{figure}[t]
	\begin{center}
		\medskip
		\includegraphics[width=.9\linewidth]{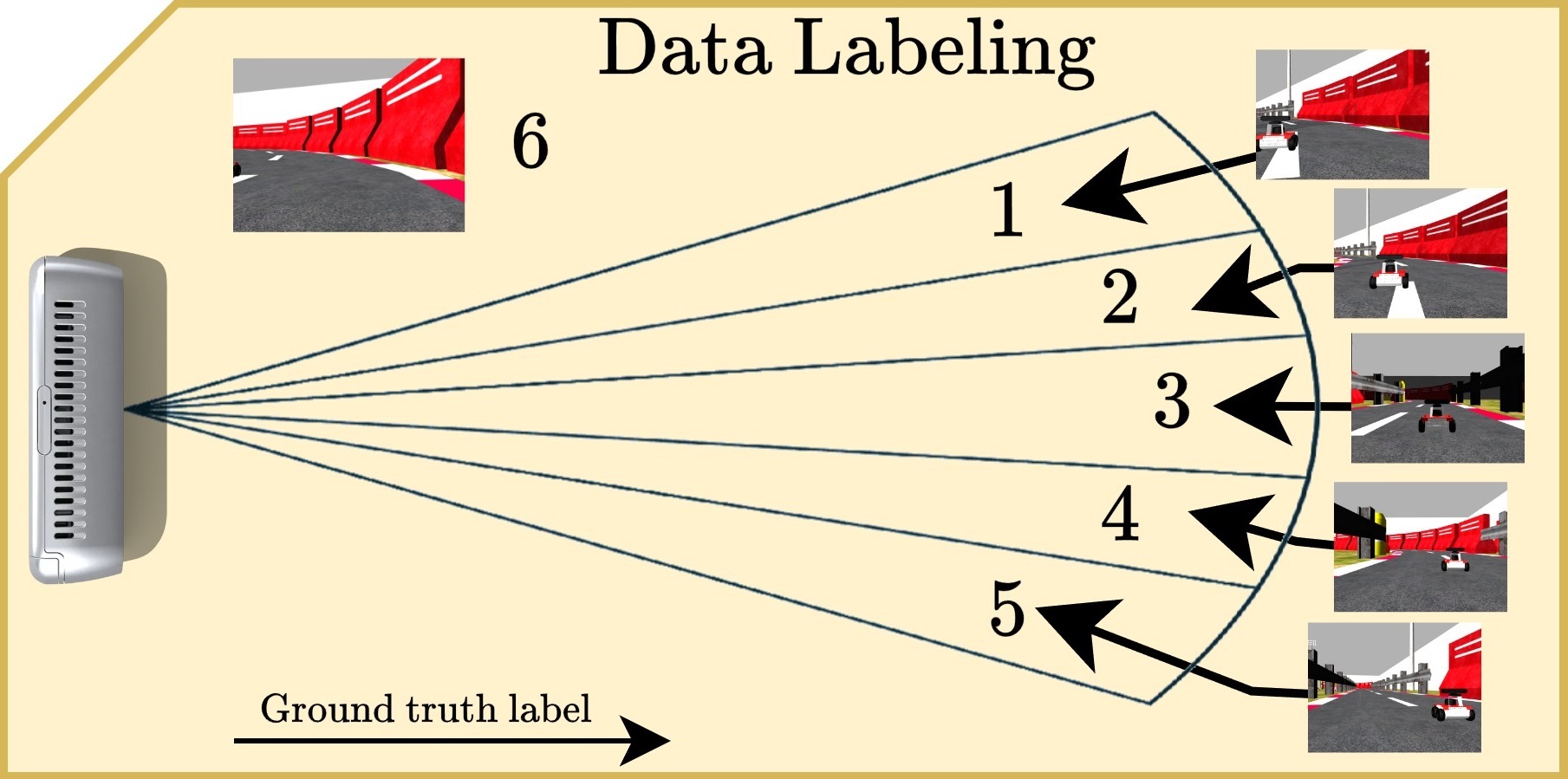}
		\caption{{Data Labeling}. Image data labeling example with six classes.
			\label{DataEncoding}}
	\end{center}
	\vspace{-4ex}
\end{figure}
This section discusses the methods to estimate the angle $\phi_i$ and distance $L_i$ using the follower's onboard camera.  
\subsection{Estimating $\phi_i$ Using a Temporal Filter-Based DNN}
To estimate the follower's bearing angle $\phi_i$ as illustrated in Fig.~\ref{fig:lead-fol}, this subsection presents a DNN-based estimator combined with a linear temporal filter, which maps real-time image data to angle estimates. The estimation task is framed as an image classification problem. The following subsections detail the processes of data collection, model training, and inference. 

\begin{figure}[t]
	\begin{center}
		\medskip
		\includegraphics[width=\linewidth]{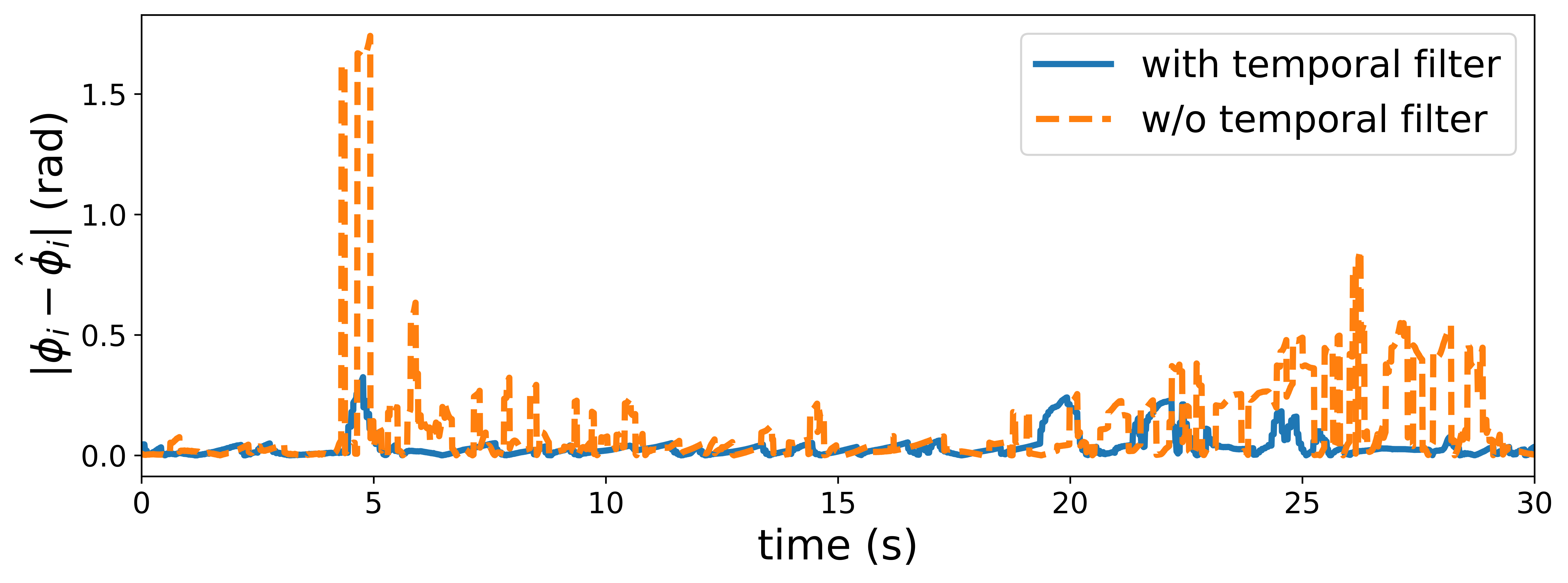}
		\caption{{Comparison Results of Angle Estimation Error with and without Linear Temporal Filter}. 
			\label{DNNComp}}
	\end{center}
	\vspace{-4ex}
\end{figure}

\begin{figure}[t]
	\begin{center}
		\medskip
		\subcaptionbox{RGB Image}{\includegraphics[width=0.48\linewidth]{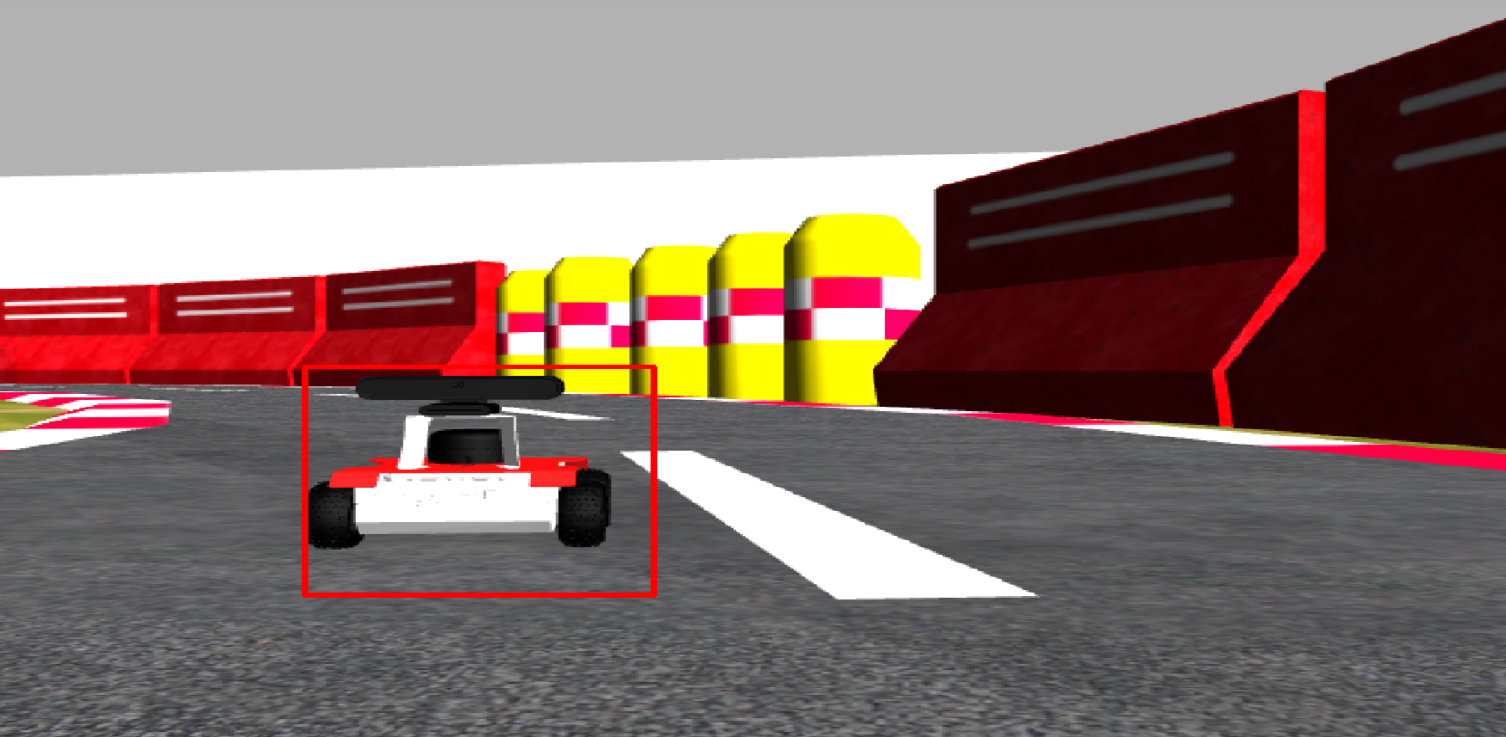}} 
		\subcaptionbox{Depth Image}{\includegraphics[width=0.48\linewidth]{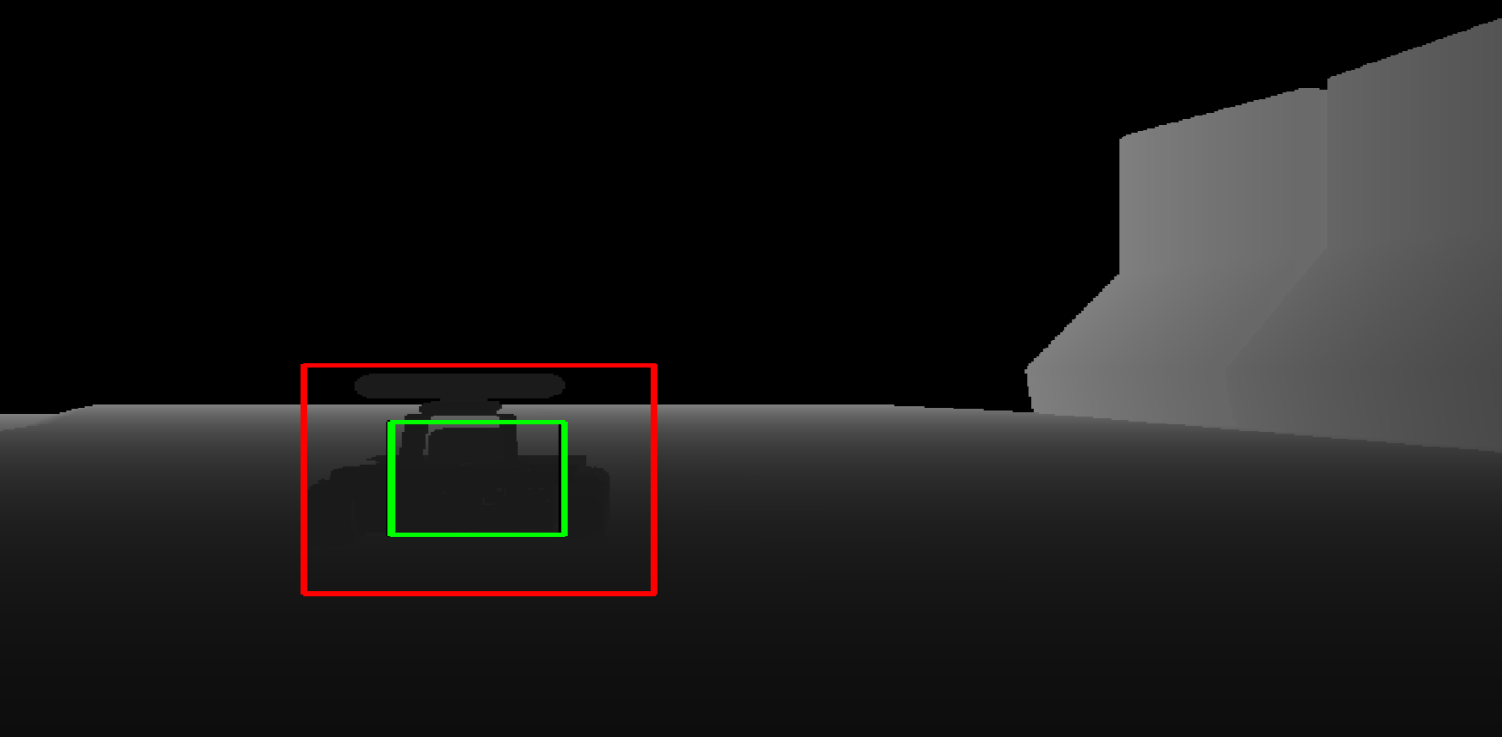}}
		\caption{{Double Bounding Box-based Estimator.}
			\label{DistanceEstimation}}
	\end{center}
	\vspace{-4ex}
\end{figure}

\subsubsection{Training Data Collection}
Data collection for leader-follower system in real-world settings is challenging due to the time-consuming, costly nature and the constraints and risks associated with physical environments. This paper leverages the Gazebo simulator to generate diverse training data that capture a wide range of operating conditions for leader-follower systems under controlled, repeatable, and various scenarios. Each training data point associates an image from the follower's onboard camera with its ground truth angle. To improve the model's generalization capabilities and reduce biases from a limited training environment, we further preprocess the images using domain randomization (DR) technique \cite{tobin2017domain} to introduce various visual perturbations and variations such as:
\begin{itemize}
    \item \textbf{HSV Manipulation}: Adjusts hue, saturation, and brightness to simulate different lighting conditions.
    \item \textbf{Image Flipping}: Flips the image Horizontally to improve orientation invariance.
    \item \textbf{Blur}: Applies Gaussian, horizontal, or vertical blur to mimic motion and lens blur.
    \item \textbf{Channel Swapping}: Swaps color channels to reduce sensitivity to specific color distributions.
\end{itemize}
Experimental results in Section \ref{sec:experiment} demonstrate that the DR technique effectively generalizes model performance to environments not represented in training data. 
\subsubsection{Data Labeling}
To formulate the estimation task as a classification problem, each image in the data set must be labeled with its ground truth, which in this case is the correct angle estimate. The ground truth label is generated by partitioning the camera's FOV $2\psi_{max}$ into $n$ equal intervals with each corresponding to a class. Images not containing the leader are assigned the label $n+1$, indicating the leader is outside the FOV. The ground truth angle estimate for each class is the center of the interval corresponding to that class. Fig.~\ref{DataEncoding} shows an example of label generation with $n=5$. The resolution of the angel estimate improves exponentially as the number of classes increases. 

\begin{figure}[t] 
\begin{center}
\medskip
    \subcaptionbox{Enviorment \#1:Bird view}{\includegraphics[width=0.48\linewidth]{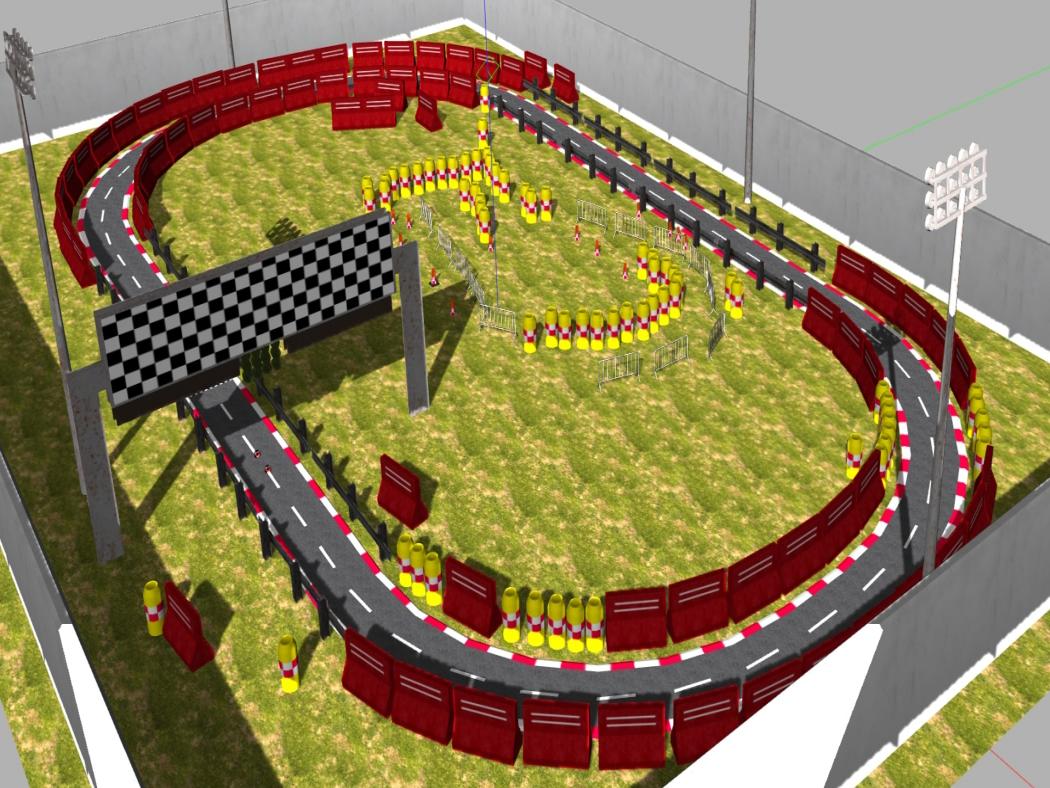}} 
    \subcaptionbox{Enviorment \#1:POV}{\includegraphics[width=0.48\linewidth]{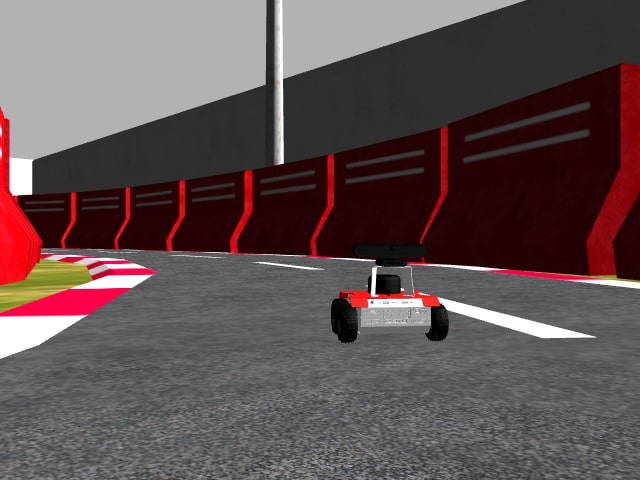}}
    \subcaptionbox{Enviorment \#2:Bird view}{\includegraphics[width=0.48\linewidth]{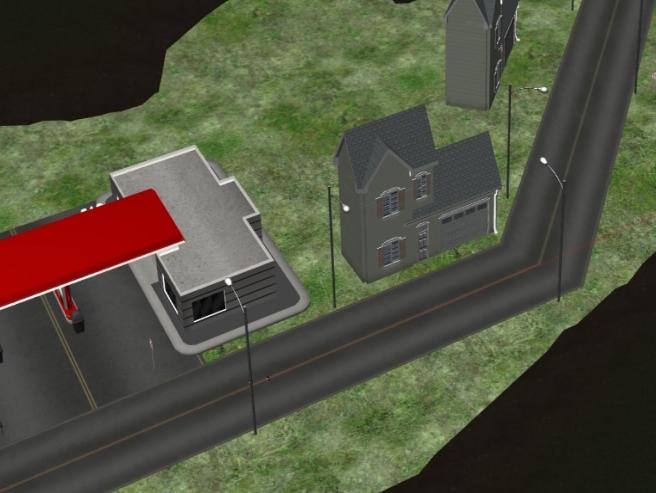}} 
    \subcaptionbox{Enviorment \#2:POV}{\includegraphics[width=0.48\linewidth]{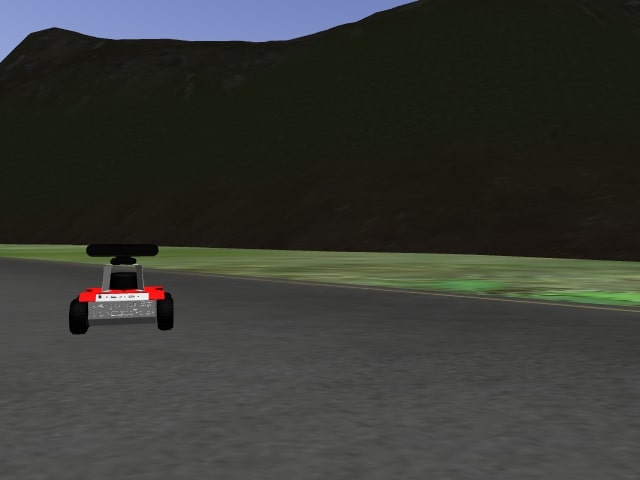}}
    \caption{Bird View and Point of View of Two Simulation Environments.}
\label{Enviorments}
\end{center}
\vspace{-4ex}
\end{figure}

\subsubsection{DNN with a Linear Temporal Filter}
The structure of the angle $\phi_i$ estimator is a Multilayer Perceptron~(MLP) combined with a linear temporal filter. This paper selects MLP over CNN for its faster inference speed and better performance in capturing the relationship between images and angles. The spatial invariance property of CNNs can lead to inaccurate angle estimates, as varying leader positions in the image result in different angle measurements. One limitation of using an MLP for angle estimation in a leader-follower system is its inability to capture temporal correlations among images, which are crucial in dynamical environments. To address this, this paper introduces a simple linear temporal filter to refine the angle estimates produced by the MLP model. Let $\hat{\phi}_{i}(t) \in \mathbb{R}$ denote the angle estimate from the MLP at time $t \in \mathbb{R}_{\geq 0}$, the linear temporal filter is defined as follows, 
\begin{align}
	\tilde{\phi}_i(t) = K_{f} \hat{\phi}_i(t) + (1 - K_{f}) \tilde{\phi}_i(t-1)
 \label{eq:filter}
\end{align}
where $\tilde{\phi}_i(t), \tilde{\phi}_i(t-1)$ are the outputs of the filter at time instants $t$ and $t-1 \in \mathbb{R}_{\geq 0}$, respectively, and $K_{f} \in (0, 1)$ is a parameter that balances the weight between the previous filtered estimate $\tilde{\phi}_i(t-1)$ and the current estimate $\hat{\phi}_i(t)$ from the MLP. The linear temporal filter improves estimation performance by ``correcting" current classification errors caused by MLP through the incorporation of prior estimates. This improvement is validated by the simulation results shown in Fig.~\ref{DNNComp} where the dashed orange line represents the absolute angle estimation error without linear temporal filter, and the blue solid line shows the error with the filter applied. The comparison clearly demonstrates that the linear temporal filter effectively reduces estimation errors by smoothing out spurious spikes from incorrect MLP predictions, thereby providing more accurate and reliable angle estimates.

\begin{figure*}[t]
	\begin{center}
		\medskip
		\includegraphics[width=0.97\linewidth]{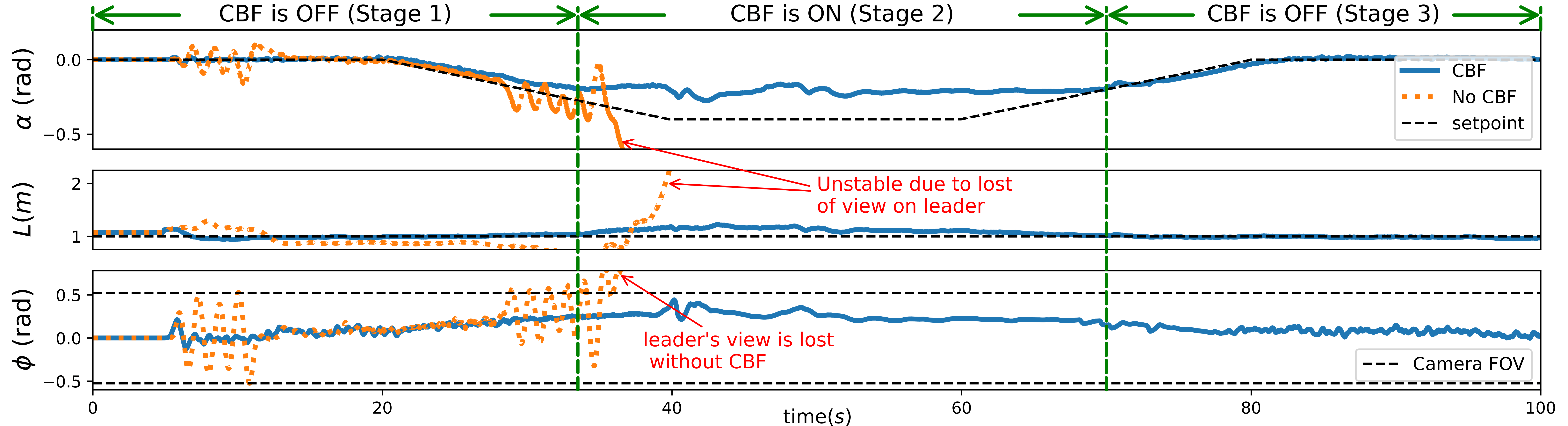}
		\caption{{Comparison of Perception-aware Safe Leader-follower Controller and Non-safe Baseline Controller in Oval Track Environment}
		\label{ValidationImg}}
	\end{center}
	\vspace{-4ex}
\end{figure*}

\subsection{Distance $L_i$ Estimation via Double Bounding Boxes}
\vspace{-0.4ex}
As shown in Fig.~\ref{fig:lead-fol}, $L_i$ represents the distance between the front of the follower and the center of the leader. Estimating the distance $L_i$ using an onboard camera is challenging due to the difficulty of accurately localizing the leader's center in the image. To address this challenge, we propose a double-bounding box method. The first, larger bounding box~(red) in Fig.~\ref{DistanceEstimation}(a) highlights the leader's general position within the image, while the second, smaller bounding box~(green) in Fig.~\ref{DistanceEstimation}(b) refines this localization to more precisely identify the leader's center.

The larger bounding box~(red) is first generated in the RGB image using the Discriminative Correlation Filter with Channel and Spatial Reliability Tracker (CSRT) \cite{lukezic2017discriminative}. The RGB image, along with this bounding box information, is then converted into depth images, where the second smaller bounding box~(green) is constructed within the larger one~(red) by reducing the width and height of the red bounding box by half. By applying sigma clipping to remove background noise from the depth pixels within the smaller bounding box, the distance $L_i$ is estimated by averaging the remaining pixel values. Similar to the angle estimation, a linear temporal filter is applied to the distance estimate $\hat{L}_i$ to reduce estimation error. Using the same notation as for the angle, let $\hat{L}_{i}$ represent the estimated distance, and $\tilde{L}_i$ denote the filtered estimate.

\section{Experiment Results}
\label{sec:experiment}

\begin{figure*}[t]
	\begin{center}
		\medskip
		\includegraphics[width=0.97\linewidth]{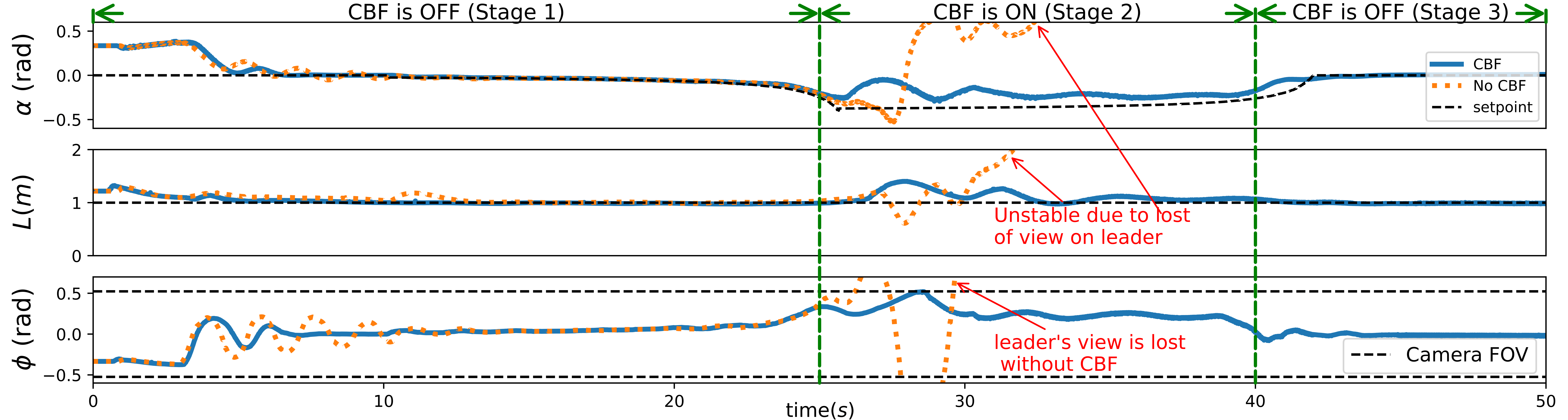}
		\caption{Comparison of Perception-aware Safe Leader-follower Controller and Non-safe Baseline Controller in Outdoor Residence Environment.
			\label{ValidationImg2}}
	\end{center}
	\vspace{-4ex}
\end{figure*}

\begin{figure*}[t]
	\begin{center}
		\medskip
		\includegraphics[width=0.97\linewidth]{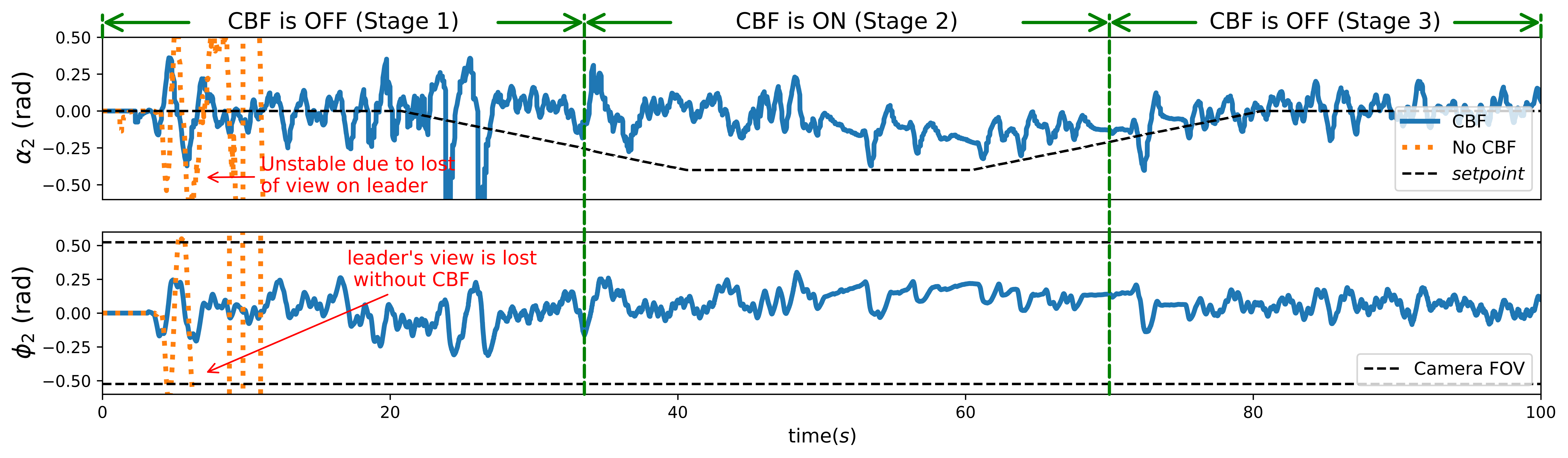}
		\caption{Comparison of Perception-aware Safe Leader-follower Controller and Non-safe Baseline Controller in Oval Track Environment for Three Robots.
			\label{ValidationImg3}}
	\end{center}
	\vspace{-4ex}
\end{figure*}

This section presents experiment results from the Gazebo simulator to verify the proposed perception-aware distributed safe leader-follower control scheme using Rosbot Pro 2.0 testing platform. The proposed approaches are evaluated in two different environments, an oval track \cite{awsrobomaker}, shown in Fig.~\ref{Enviorments}(a), and an outdoor environment, shown in Fig.~\ref{Enviorments}(b). In both environments, the experiment results demonstrate that our methods outperform the baseline by ensuring both safety and performance guarantees. All the simulations were conducted on a desktop with an Intel i9 14900KF CPU, RTX 4090 GPU, and 64GB of RAM, running Ubuntu 20.04. The CBF-QP was numerically solved in real-time using the OSQP solver \cite{osqp}. 

 \subsubsection{Simulation Setup} The parameters of the leader-follower system are configured as follows: The distance from the robot's center to its front bumper is set to $d = 0.1$m; formation controller gains are $K_{L_i} = 1.0$, $K_{\alpha_i} = 0.15$, and the CBF gains are $\gamma_i = 0.45, \forall i\in[1, 4]$ as specified by the CBF inequality \eqref{ineq:cbf-compact}. The onboard camera parameters are set to $D_\text{min} = 0.6 $m, $D_\text{max} = 8.0$m for depth specifications, and $2\psi_\text{max} = 1.0472 $ rad for FOV to simulate the Orbbec Astra RGBD camera used in Rosbot Pro 2.0 \cite{rosbot}. 

The MLP architecture consists of three hidden layers with the number of neurons as $\{1024,512,256\}$. The input to the MLP model is a flattened vector from an RGB image with a dimension of $3\times88\times88$. The number of classes is set to $21$, corresponding to $21$ neurons in the output layer, resulting in an angle estimation resolution of $3.0$ degrees per class. The MLP model is trained on a dataset of $28,000$ images, capturing various operating conditions for leader-follower systems in the Oval Track environment~(Fig.~\ref{Enviorments}(a)). Model validation is performed using a testing dataset with $28,000$ images collected under the same conditions. The parameter for the linear temporal filter is set to $K_f = 0.55$ for both angle and distance estimators. \emph{The trained MLP model and estimators will be used in an outdoor nighttime environment~(Fig.~\ref{Enviorments}(a)) which is completely different and unseen by the model to test the proposed estimator's generalization and robustness.}
\subsubsection{Simulation results} The proposed perception aware safe leader-follower formation control scheme is evaluated and compared to a baseline that only applies the leader-follower formation controller from \eqref{eq:controller} in both the Oval Track and outdoor environments. Fig.~\ref{ValidationImg} and Fig.~\ref{ValidationImg2} show the comparison results in the Oval Track and outdoor environments respectively. The blue solid lines represent the trajectories of $\alpha$~(top plot), $L$~(middle plot), and $\phi$~(bottom) under the proposed approach, while the orange dashed lines represent the trajectories under the baseline. The black dashed lines in the top and middle plots represent the desired setpoints for $\alpha$ and $L$, and in the bottom plot, they denote the follower's FOV. 

Both simulations consist of three stages. In the first stage, the robots are moving in straight line, with the follower positioned directly behind the leader. In the second stage, the formation is changed to position the follower to the side of the leader. This formation change introduces conflicts between control and safety objectives, which the proposed approach successfully resolves but the baseline fails. In the third stage, the robots return to the straight line formation. Throughout the three stages of both simulations, the proposed perception-aware safe leader-follower control scheme adaptively deactivates the CBF-QP safety filter to achieve the desired formation when control and safety specifications are not in conflict~(Stage 1 and 3). When conflicts arise in Stage 2, the safety filter is activated to ensure safety by keeping the leader within follower's FOV. In contrast, the baseline strategy, which relies solely on the leader-follower formation controller fails to adjust appropriately and loses sight of the leader, causing leader-follower system to become unstable from Stage 2 onward. 

Fig.~\ref{ValidationImg3} shows the comparison results of the second leader-follower pair in a three-robot leader-follower system within the Oval Track environment. In such a system, the last robot is particularly susceptible to safety concerns due to propagated perturbations and disturbances from the preceding robots. As shown in Fig.~\ref{ValidationImg3}, the baseline system without the CBF safety filter failed almost immediately at Stage 1, whereas the system utilizing our proposed approach successfully maintained formations while consistently adhering to safety constraints.

Moreover, both the proposed control scheme and the baseline use the proposed estimators. Thus, simulation results in Fig.~\ref{ValidationImg}, Fig.~\ref{ValidationImg2}, and Fig.~\ref{ValidationImg3} implicitly demonstrate the generalization performance of the proposed DNN-based estimator with linear temporal filters. Despite being trained solely in the Oval Track environment, the estimator was applied directly to a completely different outdoor environment, where it consistently and robustly produced accurate and reliable estimations, as demonstrated by the tracking performance in Stage 1 and 3.

\section{CONCLUSIONS}
\label{sec:conclusion}
In this paper, we have developed a distributed perception aware safe leader-follower formation control scheme that explicitly considers the FOV limits of the onboard body-fixed camera by formulating them as state constraints. A distributed CBF-QP is then solved to ensure safety, as defined by these constraints. Additionally, we introduced MLP-based and double-bounding box-based estimators with linear temporal filters to provide reliable angle and distance estimates for leader-follower systems across diverse environments. The proposed approaches were validated through experiments in the Gazebo simulator. Future work will focus on implementing these methods on a real robotic platform and quantifying the uncertainty of the proposed estimators. 

%
\bibliographystyle{IEEEtran}
\balance
\bibliography{icra2025}

\end{document}